\documentclass[12pt]{amsart}
\usepackage{amssymb}
\usepackage{latexsym}
\usepackage{graphicx}
\usepackage{subfigure}
\usepackage{tikz}
\usepackage[linktocpage=true]{hyperref}
\usepackage{ulem}
\usepackage{mathrsfs}

\usepackage{color}

\hypersetup{ colorlinks   = true, %Colours links instead of ugly boxes
urlcolor  = blue, %Colour for external hyperlinks
linkcolor    = blue, %Colour of internal links
citecolor   = blue %Colour of citations
}

\def\beq{\begin{equation}}
\def\eeq{\end{equation}}

\def\bm{\begin{matrix}}
\def\em{\end{matrix}}

\newcommand{\Z}{{\mathbb Z}}
\newcommand{\R}{{\mathbb R}}

\newcommand{\T}{{\mathbb T}}

\newcommand{\N}{{\mathbb N}}

\newcommand{\ri}{{\rm i}}

\newcommand{\CH}{{\mathcal H}}
\newcommand{\CI}{{\mathcal I}}

\newcommand{\CZ}{{\mathcal Z}}

\newcommand{\CO}{{\mathcal O}}

\newcommand{\CJ}{{\mathcal J}}

\newtheorem{thm}{Theorem}[section]

\newtheorem{prop}[thm]{Proposition}
\newtheorem{lemma}[thm]{Lemma}

\theoremstyle{definition}
\newtheorem{remark}[thm]{Remark}

\newcommand{\la}{\langle}
\newcommand{\ra}{\rangle}

\definecolor{gainsboro}{rgb}{0.86, 0.86, 0.86}
\definecolor{gray-light}{rgb}{0.75, 0.75, 0.75}
\definecolor{gray}{rgb}{0.5, 0.5, 0.5}

\begin{document}

\title[Ballistic transport and decaying potential]{Ballistic Transport for Discrete Multi-Dimensional Schr\"odinger Operators With Decaying Potential}

%\title[]{Ballistic transport for Sch\"odinger operator with decaying potential}

\author{David Damanik}
\address{Department of Mathematics, Rice University, 6100 S. Main Street, Houston, Texas
77005-1892, U.S.A.}
\email{damanik@rice.edu}

\thanks{D.\ D.\ was supported in part by NSF grants DMS--2054752 and DMS--2349919}

\author{Zhiyan Zhao}
\address{Universit\'e C\^ote d'Azur, CNRS, Laboratoire J. A. Dieudonn\'{e}, 06108 Nice, France}
\email{zhiyan.zhao@univ-cotedazur.fr}
\thanks{Z. Zhao was partially supported by ANR grant for ANR-22-CE40-0016 and partially supported by NSFC grant 12471178.}


\date{\today}

\begin{abstract}

We consider the discrete Schr\"odinger operator $H = -\Delta + V$ on $\ell^2(\Z^d)$ with a decaying potential, in arbitrary lattice dimension $d\in\N^*$, where $\Delta$ is the standard discrete Laplacian and $V_n = o(|n|^{-1})$ as $|n| \to \infty$.
%We prove the absence of singular continuous spectrum for $H$.
For the unitary evolution $e^{-\ri tH}$, we prove that it exhibits ballistic transport in the sense that, for any $r > 0$, the weighted $\ell^2-$norm
$$\|e^{-\ri tH}u\|_r:=\left(\sum_{n\in\Z^d} (1+|n|^2)^{r} |(e^{-\ri tH}u)_n|^2\right)^\frac12  $$
grows at rate $\simeq  t^r$ as $t\to \infty$, provided that the initial state $u$ is in the absolutely continuous subspace and satisfies $\|u\|_r<\infty$.

The proof relies on commutator methods and Mourre estimate, which yields quantitative lower bounds on transport for operators with purely absolutely continuous spectrum over appropriate spectral intervals. 
%Compactness arguments and localized spectral projections are used to extend the result to perturbed operators, extending the classical result for the free Laplacian to a broader class of decaying potentials.

\end{abstract}

%\begin{keyword}
%%% keywords here, in the form: keyword \sep keyword
%1-d quantum harmonic oscillator; time quasi-periodic; reducibility; growth of Sobolev norms
%%% PACS codes here, in the form: \PACS code \sep code
%
%%% MSC codes here, in the form: \MSC code \sep code
%%% or \MSC[2008] code \sep code (2000 is the default)
%\MSC[2010] 	35Q40; 35Q41; 47G30
%\end{keyword}

%% \linenumbers

\maketitle

\section{Introduction and Main Result}

\subsection{General Setting and Questions}

A main motivation to perform a spectral analysis of a Schr\"odinger operator
$H = - \Delta + V$
is given by the known connections between spectral properties of $H$ and the long-term asymptotics of the associated unitary group $e^{-\ri tH}$, with which one can describe the solutions of the associated time-dependent Schr\"odinger equation. That is,
$
\psi(t) = e^{-\ri tH} \psi_0
$
is the unique solution of the time-dependent equation
$$
\ri  \partial_t \psi = H \psi, \qquad \left. \psi\right|_{t = 0} = \psi_0.
$$
A fundamental result establishing such a connection is given by the RAGE theorem; compare, for example, \cite[Theorem~1.6.7]{DF22}. More refined connections can be found in \cite{DF22, L96} and references therein.

Generally speaking, the more regular the spectral measure $\mu = \mu_{H,\psi_0}$ associated with the pair $(H, \psi_0)$ is, the greater the tendency of $\psi$ is to leave compact regions in space as time grows.

In order to be more specific, let us consider the case where the operator $H$ acts in the Hilbert space $\ell^2(\Z^d)$; analogous statements exist in the case of the continuum, $L^2(\R^d)$, or the half line, $\ell^2(\Z_+)$ or $L^2(\R_+)$.

Consider the Lebesgue decomposition of $\mu$ into its absolutely continuous, singular continuous, and pure point parts:
$$
\mu = \mu_\mathrm{ac} + \mu_\mathrm{sc} + \mu_\mathrm{pp}.
$$
Let us first discuss the question of whether the time evolution leaves compact regions or is confined to a suitable compact region, up to an arbitrarily small error. The RAGE theorem states that
\begin{eqnarray*}
\mu = \mu_\mathrm{pp} & \Longleftrightarrow& \forall \  \varepsilon > 0, \  \exists \  N,  \    \forall  \   t \in \R \; : \; \sum_{|n| > N} |\langle \delta_n, \psi(t) \rangle |^2 < \varepsilon, \\
\mu = \mu_\mathrm{sc} + \mu_\mathrm{ac} &  \Longleftrightarrow &\forall  \  N \; : \; \lim_{T \to \infty} \frac{1}{2T} \int_{-T}^T \sum_{|n| \le N} |\langle \delta_n, \psi(t) \rangle |^2 \, dt = 0, \\
\mu = \mu_\mathrm{ac} & \Longrightarrow& \forall  \  N \; : \; \lim_{|t| \to \infty} \sum_{|n| \le N} |\langle \delta_n, \psi(t) \rangle |^2 = 0.
\end{eqnarray*}
Recall that for a normalized initial state, $\| \psi_0 \| = 1$, $\sum_{|n| \le N} |\langle \delta_n, \psi(t) \rangle |^2$ gives the probability of the wavefunction being in the ball of radius $N$ centered at the origin at time $t$, so these statements indeed have the interpretation alluded to above.

We will be especially interested in cases where the spectral measure is absolutely continuous (a.c. for short).
%the value $1$ is taken, as this is a weak form of ballistic transport statement.
This is in response to the following general question: for initial states $\psi_0$ with $\mu = \mu_\mathrm{ac}$, is transport ballistic (formulated in Subsection~\ref{ss.1.2})? There are general remarks that are relevant to this question (see the survey \cite{DMY2024} for a much more in-depth discussion):

\begin{itemize}

\item[(i)] One expects this to be true for many models of interest, but there is no general result.

\item[(ii)] In fact, there cannot be a completely general result as there are counterexamples, which can be constructed via inverse spectral theory in higher dimensions; see Bellissard and Schulz-Baldes \cite{BS00}.

%\item[(iii)] On the other hand, in one dimension ($\ell^2(\Z)$ or $\ell^2(\Z_+)$), we always have $\tilde \beta^\pm(p) = 1$ for every $p > 0$ when $\mu = \mu_\mathrm{ac}$; see Last \cite{L96}. This in turn implies $\beta^+(p) = 1$ for every $p > 0$. On the other hand, whether $\beta^-(p) = 1$ will always hold under the assumption  $\mu = \mu_\mathrm{ac}$ is at present wide open. The currently available tools do not allow us to answer this question for general potentials $V$.
\end{itemize}

Whether or not  $\mu = \mu_\mathrm{ac}$ implies ballistic transport in one dimension being a difficult question in general, it is natural to explore it for specific interesting classes of potentials. This has indeed been done. For periodic $V$, it was shown by Asch and Knauf \cite{AK98} in the continuum setting and Damanik, Lukic, and Yessen \cite{DLY2015} in the discrete case. An extension to suitable limit-periodic potentials was obtained by Fillman \cite{F17} in the discrete case and by Young \cite{Y23} in the continuum. Finally, suitable quasi-periodic potentials were studied by Zhao \cite{Z2016, Z2017}, Zhang and Zhao \cite{ZZ2017}, and Ge and Kachkovskiy \cite{GK23}. In higher dimensions, there are results of Asch and Knauf \cite{AK98}, Black, Damanik, Malinovitch, and Young \cite{BDMY23}, Boutet de Monvel and Sabri \cite{BS23}, Fillman \cite{F21}, and Karpeshina, Lee, Parnovski, Shterenberg, and Stolz \cite{KLSS17, KPS21}.\footnote{It should be noted, however, that \cite{KLSS17, KPS21} consider the time-averaged evolution, whereas this paper and the other papers mentioned above do not.}

While periodic, limit-periodic, and quasi-periodic potentials are of course of great importance with respect to actual physical models, another very important class is that of decaying potentials, and the absence of a ballistic transport result for them is quite striking. The present paper seeks to address this and produce a ballistic transport result for suitable decaying potentials.
%For simplicity, we will focus on the discrete half-line case. It turns out that even in this simple setting, the proof of ballistic transport is quite involved. The specific result we obtain will be described in Subsection~\ref{ss.1.2}.
%In Subsection~\ref{ss.1.3} we will return to our general discussion and place our result in the context of general expectations, leading to explicit open problems we will state and discuss there.

\subsection{Ballistic Transport}\label{ss.1.2}
Consider the $d-$dimensional discrete Schr\"odinger operator $H_V=-\Delta + V$ defined through the discrete Laplacian
\begin{equation}\label{discreteLaplacian}
\Delta : \ell^2(\Z^d) \to \ell^2(\Z^d),\quad (\Delta u)_n=\sum_{j=1}^d (\Delta_j u)_n:=\sum_{j=1}^d (u_{n+e_j}+u_{n-e_j}),
\end{equation}
with $\{e_j\}_{j=1}^d$ the canonical basis of $\Z^d$, and the multiplication by the potential $V=(V_n)$.

%Our first main result is about the spectral type of $H_V$.

%\begin{thm}\label{thm-spec}
%If $V=(V_n)$ satisfies
%\begin{equation}\label{decayV-disc}
%V_n= o(|n|^{-1}),\quad |n|\to\infty,
%\end{equation}
%then the Schr\"odinger operator $H_V=-\Delta+V$ has no singular continuous spectrum.
%Moreover, for any closed interval $\CI\subset (-2d,2d)$, there are at most finitely many eigenvalues of $H_V$ in $\CI$ and each of these eigenvalues is of finite multiplicity.
%\end{thm}
%
%The decay condition \eqref{decayV-disc} of $V$ arises naturally from our proof.
%For a conjugate operator $A$ which establishes the Mourre estimate (see Section \ref{sec_Mourre}) for $H_V$, the condition \eqref{decayV-disc} implies that $H_V\in C^1(A)$. Under a stronger short-range decay condition of $V$, the absence of singular continuous (s.c. for short) spectrum for $H_V$ is demonstrated in \cite{BS99}. This potential ensures that $H_V$ belongs to the class $\mathscr{C}^{1,1}(A)\subset C^1(A)$ (as defined in \cite[Definition 6.2.2]{ABG1996}).
%%According to \cite[Theorem 7.4.2]{ABG1996}, for $H_V \in \mathscr{C}^{1,1}(A)$, if the Mourre estimate holds, then $H_V$ has no s.c. spectrum.
%Additionally, it is noted in \cite[3.3]{BS99} that it remains an open question whether the $C^1(A)$ regularity property is sufficient for the absence of s.c. spectrum of $H_V$ in the subset on which the Mourre estimate holds.
%See Section \ref{sec-nosc}  for more detailed discussions.
%
%
%
%
%
%

%
%
%\subsection{Ballistic Transport}\label{ss.1.2}

For $r\in\R_+$ and $d\in\N^*$, let us define the subspace $\CH^r(\Z^d)$ of $\ell^2(\Z^d)$, as
$$
\CH^r(\Z^d) := \left\{ u = (u_n)\in \ell^2(\Z^d):\sum_{n\in\Z^d} (1+|n|^2)^{r}  |u_n|^2 <\infty \right\},
$$
with $|n|:=\sqrt{n_1^2+\cdots+n_d^2}$.
%For a self-adjoint operator $H$ acting on $\ell^2(\Z^d)$,
%we are interested in the long-time behavior of its non-eigenstate, under the time evolution $e^{-\ri tH}$, in the weighted $\ell^2-$space
%$$\CH^r(\Z^d):=\left\{u=(u_n)\in \ell^2(\Z^d):\sum_{n\in\Z^d} |n|^{2r} |u_n|^2 <\infty\right\},$$
%where $r\in\R_+$ and $|n|:=\sqrt{n_1^2+\cdots+n_d^2}$.
It also naturally defines the weighted $\ell^2-$norm $\|\cdot\|_r$ on $\CH^r(\Z^d)$ via
$$
\| u \|_r := \left( \sum_{n\in\Z^d} (1+|n|^2)^{r}  |u_n|^2  \right)^{1/2}.
$$
We identify the $\|\cdot\|_{\ell^2}-$norm with the $\|\cdot\|_0-$norm.

Let $\CH_{\rm ac}(H_V)$ be the absolutely continuous spectral subspace associated with $H_V$. For $r\geq 0$, let
$\CH_{\rm ac}^r(\Z^d):=\CH^r(\Z^d)\cap \CH_{\rm ac}(H_V)$. Here is our main result:

\begin{thm}\label{thm-ballistic-decay}
If $V=(V_n)$ satisfies
\begin{equation}\label{decayV-disc}
V_n= o(|n|^{-1}),\quad |n|\to\infty,
\end{equation}
then, the unitary evolution $e^{-\ri t H_V}$ exhibits ballistic transport in the sense that, for any $u\in \CH_{\rm ac}^r(\Z^d)\setminus\{0\}$ and any $r> 0$, there exists a constant $C_{u,r}>1$ such that
\beq\label{ballistic-decayingV}
C_{u,r}^{-1}\leq \frac{\|e^{-\ri tH_V} u\|_{r}}{t^r}\leq C_{u,r} \quad \forall \ t \ge 1.
\eeq
\end{thm}

\begin{remark}
In the terminology of the survey \cite{DMY2024}, Theorem~\ref{thm-ballistic-decay} establishes ballistic transport in the norm sense for all moments under the decay assumption \eqref{decayV-disc}.
\end{remark}

The condition \eqref{decayV-disc} is known to be the sharp decay condition that ensures that the interior of the essential spectrum is purely a.c. in one dimension \cite{R1998}. Note that Wigner-von Neumann potentials satisfy $V_n= O(|n|^{-1})$, $|n|\to\infty$, and allow for eigenvalues embedded in the interior of the essential spectrum \cite{DK04, RS-4}.\footnote{See especially \cite[Example on p.52]{DK04} for how to embed an eigenvalue in the discrete case.} Whether or not \eqref{decayV-disc} ensures purely a.c. interior essential spectrum in higher dimensions as well is not known (to the best of our knowledge). 
%At this point we cannot rule out the existence of isolated eigenvalues embedded in the interior of the essential spectrum. 
It is worth pointing out that the absence of embedded eigenvalues is known in the continuum case due to a classical result of Kato under the analogous decay condition, $V(x) = o(|x|^{-1})$, \cite{K59}. However, a discrete analogue of Kato's result in dimensions $d \ge 2$ is not yet known. For work in this direction, see Mandich \cite{M19}.

Clearly, the ballistic lower bound in \eqref{ballistic-decayingV} cannot hold for eigenstates, and hence we naturally restrict our attention in Theorem~\ref{thm-ballistic-decay} to initial states in the absolutely continuous subspace.

\medskip

The remainder of the paper is organized as follows.
As a key ingredient of the proof, the Mourre estimate for $H_V$ is established in Section~\ref{sec_Mourre}.
Based on this estimate, 
%Theorem~\ref{thm-spec} is proved in Section~\ref{sec-nosc}, and 
the order$-1$ ballistic lower bound is derived in Section~\ref{sec1-lower}.
Finally, the proof of Theorem~\ref{thm-ballistic-decay} is completed in Section~\ref{sec-BT}, where ballistic upper and lower bounds of arbitrary order are established.

\section{Mourre estimate}\label{sec_Mourre}

Define the weight operator $Q:\CH^1(\Z^d)\to \ell^2(\Z^d)$ by
\begin{equation}\label{positionOp}
(Qu)_n:=\sqrt{1+|n|^2} \, u_n,\quad n\in\Z^d.
\end{equation}
We introduce the conjugate operator $$A:=-\ri[H_V,-Q^2]=\ri[Q^2,\Delta]$$ which, $u\in D(A)$,
can be written explicitly as
$$(Au)_n
%=\left(\sum_{j=1}^d A_ju\right)_n
=-\ri\sum_{j=1}^d\left((2n_j+1)u_{n+e_j}-(2n_j-1)u_{n-e_j}\right),\quad n\in \Z^d.$$
It is straightforward to verify that $\CH^1(\Z^d)\subset D(A)$ and that
$$\|Au\|_0 \le 6d\,\|u\|_1,\quad u\in \CH^1(\Z^d).$$
In particular, $D(A)$ is dense in $\ell^2(\Z^d)$.

Since the essential spectrum is invariant under compact perturbations (see \cite{RS-1}), the essential spectrum of $H_V$ coincides with that of $H_0:=-\Delta$, namely
$$
\sigma_{\rm ess}(H_V) = \sigma(H_0) = [-2d,2d].
$$
Indeed, under assumption~\eqref{decayV-disc}, the multiplication operator $V=(V_n)$ is compact on $\ell^2(\Z^d)$, since $V_n \to 0$ as $|n|\to\infty$.

A further computation shows that, for $u\in D(A)$,
%\footnote{Similar computations can be found in \cite{BS99}}
\begin{eqnarray}
\ri[H_0, A]u&=&2\sum_{j=1}^d (4 {\rm I}-\Delta_j^2)u,\label{commutators-Lap}\\
(\ri[V, A]u)_n&=& \sum_{j=1}^d\left(\left(2n_j+1\right) (V_n - V_{n+e_j}  ) u_{n+e_j}\right.\label{commutators-V}\\
& & \  \  \  \  \  \  \left.-\left(2n_j-1\right) (V_n - V_{n-e_j}  ) u_{n-e_j}\right).\nonumber
\end{eqnarray}
Under assumption~\eqref{decayV-disc}, the operator $\ri[V, A]$ extends to a compact operator on $\ell^2(\Z^d)$.
Consequently, $\ri[H_V, A]$ extends to a bounded operator on $\ell^2(\Z^d)$.  In the framework of Amrein, Boutet de Monvel and Georgescu, this implies that $H_V\in C^1(A)$ (see \cite[Definition 6.2.2 and Theorem 6.2.10]{ABG1996} and \cite[3.2.2]{BS99} for precise definitions and properties).

The absence of s.c. spectrum was established in \cite{BS99} under a stronger short-range decay assumption on the potential $V$, which ensures that $H_V$ belongs to the class $\mathscr{C}^{1,1}(A)\subset C^1(A)$ (see \cite[Definition 6.2.2]{ABG1996}). According to \cite[Theorem 7.4.2]{ABG1996}, if $H_V \in \mathscr{C}^{1,1}(A)$ and the Mourre estimate \eqref{Mourre-alpha} holds, then $H_V$ has no s.c. spectrum. It is further pointed out in \cite[Section 3.3]{BS99} that it remains an open question whether the weaker regularity assumption $C^1(A)$ suffices to ensure the absence of singular continuous spectrum for $H_V$ on the set where \eqref{Mourre-alpha} holds; compare also the discussion in \cite{GM18}.

\medskip

Define the set of thresholds
$$\CZ:=[-2d,2d]\cap\{2d+4k\}_{k\in\Z}. $$
Then the set $[-2d,2d]\setminus\CZ$ consists of $d$ disjoint open intervals, each of length $4$.
We denote these intervals by $\CJ_j$, $1\le j \le d$.

\begin{lemma}\label{lemMouEsti}
Let $E\in [-2d,2d]\setminus\CZ$. Then there exists $\delta_0>0$ such that, for $\CI_0=(E-\delta_0,E+\delta_0)$, there exists $0<\alpha<8d^{-2}$ satisfying
\begin{equation}\label{Mourre-alpha}
\chi_{\CI_0}(H_V)\ri[H_V,A]\chi_{\CI_0}(H_V)
\ge \alpha\,\chi_{\CI_0}(H_V)
+ \chi_{\CI_0}(H_V) K_V \chi_{\CI_0}(H_V),
\end{equation}
where $K_V$ is a compact operator on $\ell^2(\Z^d)$.
\end{lemma}

\begin{remark} According to \cite{M1981}, the estimate \eqref{Mourre-alpha} implies that, for any interval $I\subset [-2d,2d]\setminus\CZ$, $H_V$ has at most finitely many eigenvalues in $I$, each of finite multiplicity.

%Since $H_V \in C^1(A)$, assertion (i) of Theorem~\ref{thmLAP} follows from \cite[Corollary 7.2.11]{ABG1996}. The remainder of this section is devoted to the proof of assertion (ii) of Theorem~\ref{thmLAP}.
\end{remark}

\begin{remark}
Beyond the setting of $\ell^2(\Z^d)$, Mourre estimates have also been employed to establish the absence of s.c. spectrum for adjacency operators on graphs; see, for instance, \cite{MRT07}. Moreover, Mourre theory plays a central role in scattering theory for Schr\"odinger operators as well as for a broader class of dispersive PDEs; see, e.g., \cite{BS12, DG97, GN98, GNRS20, M2022, MM2026, PSS81}.
\end{remark}

\proof
Let $E\in \CJ_j\subset [-2d,2d]\setminus\CZ$ for some $1\leq j\leq d$.
Then a sufficiently small neighborhood of $E$ is still contained in $\CJ_j$.
Moreover, there exists $\delta_0>0$ such that
\begin{equation}\label{distIZ}
{\rm dist}((E-2\delta_0, E +2\delta_0), \CZ)> \frac{d\sqrt{\alpha}}{\sqrt{2}} \ {\rm for \ some } \ 0<\alpha<8d^{-2}.
 \end{equation}

Under the Fourier transform
$$\widehat{u}(\theta)=\sum_{n\in\Z^d} u_n e^{\ri \la n, \theta\ra}, \quad \theta\in \T^d=[0,2\pi)^d,$$
the operator $H_0=-\Delta$ becomes a multiplication operator:
$$\widehat{(-\Delta u)}(\theta)
= -\left( \sum_{j=1}^d \left(e^{\ri \theta_j}+e^{-\ri\theta_j}\right) \right)\widehat{u}(\theta)
=- \left(2\sum_{j=1}^d \cos(\theta_j) \right)\widehat{u}(\theta).
$$
The thresholds in $\CZ$ correspond to the values
$$E(\theta):=- 2\sum_{j=1}^d \cos(\theta_j) \quad {\rm with} \  \theta_j\in \{0,\pi\},$$
For $\theta$ such that $E(\theta)\in \CI_0':=(E-2\delta_0,E+2\delta_0)$ and \eqref{distIZ} holds,
there exists $1\leq j_*\leq d$ such that
$$
|\cos(\theta_{j_*})+1|,\ |\cos(\theta_{j_*})-1|
> \frac{\sqrt{\alpha}}{2\sqrt{2}}.
$$
In view of \eqref{commutators-Lap}, we have
 $$\widehat{(\ri[H_0,A] u)}(\theta)= \left(8 \sum_{j=1}^d \sin^2 (\theta_j)\right)\widehat{u}(\theta),$$
which implies
$$ \chi_{\CI_0'}(H_0)\ri[H_0,A]\chi_{\CI_0'}(H_0)\geq 8\left( \frac{\sqrt\alpha}{2\sqrt2}\right)^2\chi_{\CI_0'}(H_0)= \alpha \chi_{\CI_0'}(H_0).$$

Let $g_{\CI_0'}\in C_c^\infty(\R, [0,1])$ be supported on $\CI_0'$ and identically $1$ on $\CI_0=(E-\delta_0, E+\delta_0)$.
Noting that
$$g_{\CI_0'}(H_0)=g_{\CI_0'}(H_0)\chi_{\CI_0'}(H_0)=\chi_{\CI_0'}(H_0)g_{\CI_0'}(H_0),$$
and therefore
$$ g_{\CI_0'}(H_0)\ri[H_0,A]g_{\CI_0'}(H_0)\geq \alpha g^2_{\CI_0'}(H_0).$$

Since $V$ is a multiplication with $V_n\to 0$, it is compact on $\ell^2(\Z^d)$.
Hence $(H_V - z)^{-1} - (H_0 - z)^{-1}$ is compact for $z\notin\R$.
By the Helffer--Sj\"ostrand functional calculus \cite{HS1989}, it follows that
$g_{\CI_0'}(H_V)-g_{\CI_0'}(H_0)$ and $g^2_{\CI_0'}(H_V)-g^2_{\CI_0'}(H_0)$ are compact.
Moreover, under assumption~\eqref{decayV-disc}, the commutator $\ri[V,A]$ is compact.
It follows that
$$g_{\CI_0'}(H_V)\ri[H_V,A]g_{\CI_0'}(H_V)\geq \alpha g^2_{\CI_0'}(H_V)+K_V, $$
where $K_V$ is the compact operator defined by
\begin{eqnarray*}
K_V&:=&g_{\CI_0'}(H_V)\ri[V,A]g_{\CI_0'}(H_V)\label{KV}\\
& & + \,
(g_{\CI_0'}(H_V)-g_{\CI_0'}(H_0))\ri[H_0,A]g_{\CI_0'}(H_V)\nonumber\\
& & + \, g_{\CI_0'}(H_0)\ri[H_0,A](g_{\CI_0'}(H_V)-g_{\CI_0'}(H_0)) \nonumber\\
& & - \, \alpha (g^2_{\CI_0'}(H_V)-g^2_{\CI_0'}(H_0)) .\nonumber
\end{eqnarray*}
Each term in $K_V$ is compact, being a product of bounded and compact operators.
Since $g_{\CI_0'}\equiv 1$ on $\CI_0=(E-\delta_0, E+\delta_0)$, we have
\begin{eqnarray*}
& &\chi_{\CI_0}(H_V)\ri[H_V,A]\chi_{\CI_0}(H_V) \\
&=& \chi_{\CI_0}(H_V) g_{\CI_0'}(H_V)\ri[H_V,A]g_{\CI_0'}(H_V)  \chi_{\CI_0}(H_V)\\
&>&  \alpha \chi_{\CI_0}(H_V)  + \chi_{\CI_0}(H_V)  K_V  \chi_{\CI_0}(H_V),
\end{eqnarray*}
which proves \eqref{Mourre-alpha}. \qed
\begin{remark}\label{rmk_delta}
In view of the above proof, we observe that the parameter $\delta_0$ in \eqref{Mourre-alpha} depends only on the distance of $E$ to the threshold set $\CZ$ (and the choice of $\alpha$), and is independent of the potential $V$.
%{\clo Moreover, for two potentials $V$ and $W$, both satisfying the decay rate condition \eqref{decayV-disc}, we have,
%in view of \eqref{KV},
%\begin{eqnarray*}
%K_V-K_W&=&(g_{\CI_0'}(H_V)-g_{\CI'}(H_W))\ri[V,A]g_{\CI_0'}(H_V) \\
%& & + \,  g_{\CI_0'}(H_W))\ri[V-W,A]g_{\CI_0'}(H_V)\\
%& & + \, g_{\CI_0'}(H_W))\ri[W,A](g_{\CI_0'}(H_V)-g_{\CI_0'}(H_W))\\
%& & + \, (g_{\CI_0'}(H_V)-g_{\CI_0'}(H_W))\ri[H_0,A]g_{\CI_0'}(H_V)\\
%& & + \, g_{\CI_0'}(H_W)\ri[H_0,A](g_{\CI_0'}(H_V)-g_{\CI_0'}(H_W))\\
%%& & + \, g_{\CI_0'}(H_0)\ri[H_0,A](g_{\CI_0'}(H_V)-g_{\CI_0'}(H_W))\\
%& & - \, \alpha(g_{\CI_0'}(H_V)-g_{\CI_0'}(H_W)) .
%\end{eqnarray*}
%Hence, there exists $c_*>0$, depending only on $g_{\CI_0'}$ and $\alpha$, such that
%\begin{equation}\label{KVKW}
%\|K_V-K_{W}\|_0 \leq \|[V-W,A]\|_0+c_*(\|[V,A]\|+\|[W,A]\|)\|V-W\|_0,
%\end{equation}
%where $\|\cdot\|_0$ denotes the operator norm on $\ell^2(\Z^d)$.}
\end{remark}

\section{Order$-1$ Ballistic Lower Bound}\label{sec1-lower}

The order $-1$ ballistic lower bound, i.e., the lower bound in \eqref{ballistic-decayingV} with $r=1$, is established in the following proposition.
Throughout this section, the implicit constants in estimates denoted by ``$\lesssim$'' and ``$\gtrsim$'' are independent of both the time $t$ and the initial state $u$.

\begin{prop}\label{prop-mourre-ballistic}
Given $0<\alpha<8 d^{-2}$, let $I_\alpha\subset[-2d,2d]\setminus\CZ$ be a finite union of open intervals such that there exists a compact operator $K_V^\alpha$ satisfying
 \begin{equation}\label{Mourre-alpha-prop}
 \chi_{I_\alpha}(H_V)\ri[H_V,A]\chi_{I_\alpha}(H_V)> \alpha\chi_{I_\alpha}(H_V)+\chi_{I_\alpha}(H_V)K^\alpha_V\chi_{I_\alpha}(H_V).
 \end{equation}
Let $u\in \CH_{\rm ac}^1(\Z^d)$ with $\chi_{I_\alpha}(H_V) u\neq 0$. Then\begin{equation}\label{ballistic-lower}
\frac{\|e^{-\ri tH_V} u\|_{1}}{t}\gtrsim \alpha^\frac12  \|\chi_{I_\alpha}(H_V) u\|_{0},\quad  \forall \ t > 0.
\end{equation}
\end{prop}

\begin{remark}\label{rmk_Ialpha}
The existence of the intervals $I_\alpha$ such that \eqref{Mourre-alpha-prop} holds is guaranteed by Lemma \ref{lemMouEsti}. According to the proof of Lemma \ref{lemMouEsti}, it is enough to choose $I_\alpha$ so that
${\rm dist}(I_\alpha,\CZ)>2^{-\frac12}d\cdot \alpha$.
Moreover, on each component of $I_\alpha$, the compact operator $K^\alpha_V$ can be written explicitly, and its compactness follows from the compactness of both $V$ and $[V,A]$.
\end{remark}

Throughout the proof of Proposition \ref{prop-mourre-ballistic}, we define
$B:=\ri[H_V,A]$,
where $A$ is the conjugate operator given by $A=-\ri[H_V,-Q^2]$ in the Mourre estimate.
For simplicity, we omit the subscript $V$ in $H_V$ since the potential $V$ is fixed.

Following the arguments of \cite[Proposition~1.2]{CH1992} and \cite[Proposition~4.6]{DMY2024}, it suffices to establish a quadratic-in-time lower bound for
\begin{eqnarray}
\|Q e^{-\ri tH}u\|_0^2&=&\la   u, e^{\ri tH} Q^2 e^{-\ri tH}    u \ra\nonumber \\
&=& \|Q u \|_0^2 + t \la  u, A   u  \ra
 +  \int_0^t \int_0^s   \left\la  e^{-\ri \tau H}  u, B e^{-\ri \tau H}  u \right\ra \, d\tau \, ds.\label{intdouble}
\end{eqnarray}
Since $u\in \CH^1(\Z^d)$ and $A:\CH^1(\Z^d)\to \ell^2(\Z^d)$ is bounded, we have
$$ \left|\|Q u \|_0^2 + t \la  u, A   u  \ra \|_0\right|\lesssim \|u \|_1^2+ t \|u \|_0 \|Au\|_0\lesssim \|u \|_1^2+ t \|u \|_0 \|u\|_1.$$
Therefore, it suffices to prove a quadratic lower bound in $t$ for the double integral term in \eqref{intdouble}.

Let $I_\alpha^c := [-2d,2d]\setminus I_\alpha$.
The scalar product in the double integral in \eqref{intdouble} can be
decomposed as
$$
\la e^{-\ri \tau H}u,
B e^{-\ri \tau H}u\ra
= P_{00}(\tau)+P_{10}(\tau)+P_{01}(\tau)+P_{11}(\tau),
$$
with the terms defined as
\begin{eqnarray*}
P_{00}(\tau)&:=&\left\la  e^{-\ri \tau H} \chi_{I_\alpha}(H) u, B e^{-\ri \tau H}  \chi_{I_\alpha}(H) u \right\ra,\\
P_{01}(\tau)&:=& \left\la  e^{-\ri \tau H} \chi_{I_\alpha}(H) u, B e^{-\ri \tau H}  \chi_{I_\alpha^c}(H) u \right\ra,\\
P_{10}(\tau)&:=& \left\la  e^{-\ri \tau H} \chi_{I_\alpha^c}(H) u, B e^{-\ri \tau H}  \chi_{I_\alpha}(H) u \right\ra,\\
P_{11}(\tau)&:=& \left\la  e^{-\ri \tau H} \chi_{I_\alpha^c}(H) u, B e^{-\ri \tau H}  \chi_{I_\alpha^c}(H) u \right\ra.
\end{eqnarray*}
The proof of Proposition \ref{prop-mourre-ballistic} relies on the following lemmas.

\begin{lemma}\label{lem_orth-H1}

Let $I,J\subset [-2d,2d]$ be two finite unions of intervals such that
$I\cap J=\varnothing$.
Then, for any $u\in \CH_{\rm ac}^0(\Z^d)$,
\begin{equation}\label{orth-H1}
\la e^{-\ri \tau H}\chi_I(H)u,
B e^{-\ri \tau H}\chi_J(H)u\ra
\longrightarrow 0,
\qquad \tau\to\infty .
\end{equation}
\end{lemma}
\proof Let $E_{H}(\cdot)$ be the projection-valued measure of $H$ and let $\{\delta_n\}_{n\in\Z^d}$ be the standard basis of $\ell^2(\Z^d)$.
For each $n\in\Z^d$, define the complex measure
$\mu_n(\cdot):=\la \delta_n,E_H(\cdot)u\ra$.
Since $u\in \CH_{\rm ac}^0(\Z^d)$, each $\mu_n$ is a.c. on $\sigma_{\rm ac}(H)$.
We then have
\begin{eqnarray*}
(e^{-\ri \tau H}\chi_I(H) u)_n&=&\int_{I}e^{-\ri \tau\lambda} \, d\mu_n(\lambda),  \\
(B e^{-\ri \tau H}\chi_{J}(H) u)_n&=&\sum_{m\in\Z^d} B_{nm}\int_{J}e^{-\ri \tau\lambda} \, d\mu_m(\lambda),
\end{eqnarray*}
where $B_{nm}$ denote the matrix elements of
$B=\ri[H,A]$.
By \eqref{commutators-Lap} and \eqref{commutators-V}, we have the explicit expression of $B_{nm}$.
%$$B_{nm} =\left\{\begin{array}{cl}
%8d,& m=n\\
%\pm(2n_j\pm 1)(V_n-V_{n\pm e_j}),& m=n\pm e_j\\
%-4& m=n\pm 2e_j\\
%0& {\rm otherwise}
%\end{array}
%\right. ,$$
%we see that $B$ is bounded.

Define the restricted measures
$\mu_n^I := \mu_n|_I$ and $\mu_m^J := \mu_m|_J$.
Then the total variation norms satisfy
$$\|\mu_n^I\|_{\rm TV} = |(\chi_I(H)u)_n|,\quad  \|\mu_m^J\|_{\rm TV} = |(\chi_J(H)u)_m|.$$
Introduce the operator $|B|$ with matrix elements $|B|_{nm}:=|B_{nm}|$, and the $\ell^2-$vectors
$$|\chi_I(H)u|:=(|(\chi_I(H)u)_n|)_{n\in\Z^d},\quad  |\chi_J(H)u|:=(|(\chi_J(H)u)_m|)_{m\in\Z^d}.$$
Then we obtain
\begin{eqnarray}
& &\left|\iint_{I\times J}\sum_{n,m\in\Z^d}B_{nm}\,d\mu_n(\lambda_1)\,d\overline{\mu_m}(\lambda_2)\right|\nonumber\\
&\leq&\sum_{n,m\in\Z^d}
|B_{nm}|\,\|\mu_n^I\|_{\rm TV}\|\mu_m^J\|_{\rm TV}\nonumber\\
&=&\sum_{n,m\in\Z^d}
|B_{nm}||(\chi_I(H)u)_n| |(\chi_J(H)u)_m|\nonumber\\
&=&\la|\chi_I(H)u|, |B||\chi_J(H)u| \ra\nonumber\\
&\leq&\||B|\|_{\rm op}\|\chi_I(H)u\|_0
\|\chi_J(H)u\|_0.\label{sum_measure}
\end{eqnarray}
In particular, the above bound ensures absolute convergence.
Therefore, by Fubini's theorem,
\begin{eqnarray*}
& &\left\la  e^{-\ri \tau H} \chi_{I}(H) u, B e^{-\ri \tau H}  \chi_{J}(H) u \right\ra\\
&=& \sum_{n\in\Z^d}\sum_{m\in\Z^d} B_{nm}\left(\int_{I}e^{-\ri \tau\lambda_1} \, d\mu_n(\lambda_1)\right)\overline{\left(\int_{J} e^{-\ri \tau\lambda_2} \, d\mu_m(\lambda_2)\right)}\\
&=&\iint_{I\times J}
    e^{-\ri \tau(\lambda_1-\lambda_2)}
    \left( \sum_{n,m\in\Z^d}
           B_{nm}\,d\mu_n(\lambda_1)\,d\overline{\mu_m}(\lambda_2)
    \right).
    \end{eqnarray*}
Define a finite complex measure $\nu$ on $\R$ by
$$
\nu(S)
:= \sum_{n,m\in\Z^d} B_{nm}
\int_{I\times J}
\chi_{\{\lambda_1-\lambda_2\in S\}}(\lambda_1,\lambda_2)\,
d\mu_n(\lambda_1)\, d\overline{\mu_m}(\lambda_2),
\quad S\subset \R.
$$
Then, by \eqref{sum_measure}, $|\nu|(\R)<\infty$, and we obtain
$$
\la e^{-\ri \tau H}\chi_I(H)u,\,
B e^{-\ri \tau H}\chi_J(H)u \ra
= \int_{\R} e^{-\ri \tau E}\, d\nu(E).
$$
Since $I\cap J=\varnothing$, the support of $\nu$ does not contain $0$.
By the Riemann--Lebesgue lemma, it follows that
$$
\int_{\R} e^{-\ri \tau E}\, d\nu(E) \longrightarrow 0,
\qquad \tau\to\infty,
$$
which proves \eqref{orth-H1}.\qed

%If $I\cap J\neq\varnothing$, the phase $e^{-\ri\tau(\lambda_1-\lambda_2)}$ admits a stationary region $\lambda_1=\lambda_2$ inside $I\times J$. As a consequence, the oscillatory integral may contain a non-decaying contribution, even though the diagonal set $\{\lambda_1=\lambda_2\}$ has zero measure.

\begin{lemma}\label{lem_posit-Mourre}
Given $I\subset [-2d,2d]$ a finite union of intervals with
\begin{equation}\label{Mourre}
 \chi_I(H) B  \chi_I(H) \geq  a \chi_I(H)+K,
\end{equation}
for some $a>0$ and a compact operator $K$, we have, for $u\in \CH_{\rm ac}^0(\Z^d)$, for $\tau>0$ sufficiently large,
\begin{equation}\label{posit-Mourre}
\left\la  e^{-\ri \tau H} \chi_{I}(H) u, B e^{-\ri \tau H}  \chi_{I}(H) u \right\ra\geq \frac{a}{4}\|\chi_{I}(H) u\|_0^2.
\end{equation}
\end{lemma}
\proof We can use the compactness of $K$ to obtain a strict Mourre estimate based on \eqref{Mourre} by shrinking $I$.
For $E \in I  \setminus \sigma_{\rm pp}\left(H\right)$, there exists
\begin{itemize}
  \item $\delta>0$ sufficiently small, through which we define $$\CI'=(E- 2\delta, E + 2\delta)\subset I\setminus\sigma_{\rm pp}(H),\quad \CI=(E- \delta, E + \delta),$$
  \item $g\in C_c^\infty(\R,[0,1])$,
supported on $\CI'$ and identically $1$ on $\CI$,
\end{itemize}
such that
$$
\pm g(H)  K  g(H)\leq  \frac{\alpha}{2}g^2(H),\quad g(H)  \ri[H,A] g(H)> \frac{\alpha}{2}g^2(H),
$$
which implies
\begin{equation}\label{lowerbound-E}
\chi_{\CI}(H)  \ri[H,A] \chi_{\CI}(H)> \frac{\alpha}{2}\chi_{\CI}(H).\end{equation}

Let us choose finitely many pairwise disjoint intervals
$$\CI_{j}\subset I\setminus\sigma_{\rm pp}(H), \quad 1\le j\le N,$$
such that the estimate \eqref{lowerbound-E} holds for each $\CI_{j}=(E_j-\delta,E_j+\delta)$, and
\begin{equation}\label{finite-decomp}
\left\|\chi_I(H)u-\sum_{1\leq j\leq N} \chi_{\CI_{j}}(H)u\right\|_0\le \epsilon,
\end{equation}
for some arbitrarily small $\epsilon>0$.
Then we can write
\begin{eqnarray}
& &\left\la  e^{-\ri \tau H} \chi_{I}(H) u, B e^{-\ri \tau H}  \chi_{I}(H) u \right\ra\nonumber\\&=&\sum_{1\leq j\leq N}\left\la  e^{-\ri \tau H} \chi_{\CI_j}\left(H\right) u, B e^{-\ri \tau H} \chi_{\CI_j}\left(H\right) u \right\ra\label{P00-1}\\
& & + \, \sum_{1\leq j, l\leq N\atop{j\neq l}}\left\la  e^{-\ri \tau H} \chi_{\CI_j}\left(H\right) u, B e^{-\ri \tau H} \chi_{\CI_l}\left(H\right) u \right\ra+ \CO(\epsilon), \label{P00-2}
\end{eqnarray}
where the remainder term $\CO(\epsilon)$ is uniform in $\tau$.
By Lemma~\ref{lem_orth-H1}, the sum in~\eqref{P00-2} converges to $0$ as
\(\tau\to\infty\).
On the other hand, applying \eqref{lowerbound-E} with $\CI=\CI_j$, we have
\begin{eqnarray*}
\sum_{1\leq j\leq N}\left\la  e^{-\ri \tau H} \chi_{\CI_j}\left(H\right) u, B e^{-\ri \tau H} \chi_{\CI_j}\left(H\right) u \right\ra
&\geq&  \frac{a}{2}  \sum_{1\leq j\leq N}\|\chi_{\CI_j}\left(H\right) u \|^2_0 \\
&\geq&  \frac{a}{2} \|\chi_{I}\left(H\right) u \|^2_0-\CO(\epsilon).\end{eqnarray*}
Since $\epsilon>0$ is arbitrarily small, the lower bound \eqref{posit-Mourre} follows.\qed

\medskip

\noindent{\it Proof of Proposition \ref{prop-mourre-ballistic}.}
By Lemma \ref{lem_orth-H1}, we have
\begin{equation}\label{decay-crossing}
P_{01}(\tau), \  P_{10}(\tau)\to 0,\quad  \tau\to\infty.
\end{equation}
Moreover, applying Lemma \ref{lem_posit-Mourre} together with the Mourre
estimate \eqref{Mourre-alpha-prop}, we obtain that, for $\tau>0$ sufficiently large,
\begin{equation}\label{postiP00}
P_{00}(\tau)\geq \frac{\alpha}{4} \|\chi_{I_\alpha}(H) u\|_0^2.\end{equation}

For $0<\epsilon<\alpha$ sufficiently small, let $I_\epsilon$ be the union of intervals such that, for a compact operator $K^\epsilon_V$,
 \begin{equation}\label{Mourre-epsilon-prop}
 \chi_{I_\epsilon}(H)\ri[H,A]\chi_{I_\epsilon}(H)> \epsilon\chi_{I_\epsilon}(H)+\chi_{I_\epsilon}(H)K^\epsilon_V\chi_{I_\epsilon}(H).
 \end{equation}
Since
$I_\alpha^c=(I_\epsilon\setminus I_\alpha)\cup I_\epsilon^c$,
we decompose $P_{11}(\tau)$ as
\begin{eqnarray}
P_{11}(\tau)&=&\left\la  e^{-\ri \tau H} \chi_{I_\epsilon \setminus I_\alpha}(H) u, B e^{-\ri \tau H}  \chi_{I_\epsilon \setminus I_\alpha}(H) u \right\ra\label{P11-1}\\
& &+ \, \left\la  e^{-\ri \tau H} \chi_{I_\epsilon \setminus I_\alpha}(H) u, B e^{-\ri \tau H}  \chi_{I_\epsilon^c}(H) u \right\ra\label{P11-2}\\
& &+ \, \left\la  e^{-\ri \tau H} \chi_{I_\epsilon^c}(H) u, B e^{-\ri \tau H}  \chi_{I_\epsilon \setminus I_\alpha}(H) u \right\ra\label{P11-3}\\
& &+ \, \left\la  e^{-\ri \tau H} \chi_{I_\epsilon^c}(H) u, B e^{-\ri \tau H}  \chi_{I_\epsilon^c}(H) u \right\ra\label{P11-4}.
\end{eqnarray}
By Lemma \ref{lem_posit-Mourre}, the Mourre estimate on $I_\epsilon \setminus I_\alpha\subset I_\epsilon$ implies that the term \eqref{P11-1} is non-negative.
By Lemma \ref{lem_orth-H1}, the cross terms \eqref{P11-2} and \eqref{P11-3}
tend to $0$ as $\tau\to\infty$.
In view of Remark \ref{rmk_Ialpha}, we have
$$I_\epsilon^c=[-2d,2d]\setminus I_\epsilon \subset\{E\in[-2d,2d]:{\rm dist}(E,\CZ)\leq\epsilon\}.$$
Then $\chi_{I_\epsilon^c}(H)$ converges to $0$ strongly as $\epsilon\to 0$. In particular, for $\epsilon$ sufficietly small,
$$\|\chi_{I_\epsilon^c}(H)u\|_0\leq\frac{\alpha^\frac12}{5(1+\|B\|)^\frac12} \|\chi_{I_\alpha}(H)u\|_0,$$
and consequently, the term \eqref{P11-4} is bounded by $\frac\alpha{20}\|\chi_{I_\alpha}(H)u\|^2_0$.
Therefore, we conclude that for $\tau$ large enough,
\begin{equation}\label{smallP11}
P_{11}(\tau)\ge -\frac{\alpha}{10} \|\chi_{I_\alpha}(H)u\|_0^2 .\end{equation}

Combining \eqref{decay-crossing}, \eqref{postiP00} and \eqref{smallP11}, we
obtain, for $\tau$ sufficiently large,
$$\la e^{-\ri \tau H}u,Be^{-\ri \tau H}u\ra
\geq
\frac{\alpha}{8}
\|\chi_{I_\alpha}(H)u\|_0^2 .
$$
Consequently, the double integral in \eqref{intdouble} exhibits a
$t^2-$growth, which yields the ballistic lower bound
\eqref{ballistic-lower}.\qed

\section{Proof of Theorem \ref{thm-ballistic-decay}}\label{sec-BT}

In this section, we comlete the proof of Theorem \ref{thm-ballistic-decay}. We will split it into two parts, presented in separate subsections, one addressing the upper bound and one addressing the lower bound in \eqref{ballistic-decayingV}.

\subsection{Ballistic Upper Bound}

By general principles, the upper bound in \eqref{ballistic-decayingV} holds for every $r>0$ and for any self-adjoint discrete Schr\"odinger operator.
Throughout this subsection, we only assume that $V$ is real-valued, that is, $V:\Z^d\to\R$, and we do not require that $V$ satisfy the decay condition \eqref{decayV-disc}.

%in which case the associated Schr\"odinger operator $H_V=-\Delta+V$ is self-adjoint on the domain
%$$
%D(H_V) = D(V) = \{ u \in \ell^2(\Z^d) : Vu \in \ell^2(\Z^d) \}.
%$$

Such a result is well known; compare, for example, the discussions in \cite[Appendix~B]{AW12}, \cite[Lemma~B.2]{BDMY23}, \cite[Theorem~A.1]{BS23}, \cite[Theorem 2.6.2]{DF22}, \cite[Theorem 2.22]{DT2010}, and \cite[Theorem~3]{ZZ2017}.\footnote{There is also a ballistic upper bound result in the continuum case due to Radin and Simon \cite{RS78}.} For the reader’s convenience, we include a proof below. In fact, in doing so we establish a particular instance of the upper bound, \eqref{balistic-upper-2}, that is crucial in our discussion of the order$-r$ ballistic lower bound for $0 < r < 1$ below; compare Lemma~\ref{lemSmallorder}.

\begin{prop}\label{prop-upper}
For any self-adjoint discrete Schr\"odinger operator $H_V=-\Delta+V$ on $\ell^2(\Z^d)$, we have the ballistic upper bound: there exists a constant $c_r>0$, independent of $u$, such that
\begin{equation}\label{balistic-upper-r}
\|e^{-\ri t H_V} u \|_r\leq c_r\|u\|_r t^r ,\quad \forall \ u\in\CH^r(\Z^d), \quad t \ge 1.
\end{equation}
Moreover, there exists $\tilde c_2>0$ such that
\begin{equation}\label{balistic-upper-2}\|e^{-\ri t H_V} u \|_2\leq \|u\|_2+ \|u\|_1 t+ \tilde c_2  \|u\|_0 t^2, \quad \forall \ u\in\CH^2(\Z^d), \; t \ge 1.
\end{equation}
\end{prop}
%\begin{remark} The upper bound (\ref{balistic-upper-2}) with refined coefficients will be used to establish a density argument in order to show the order$-r$ ballistic lower bound with $0<r<1$. See the proof of Lemma \ref{lemSmallorder}.
%\end{remark}
\proof Given $m\in\N^*$, for $u\in \CH^{m}(\Z^d)$, we have
\begin{equation}\label{com-int-upper}
e^{\ri tH_V}Q^m e^{-\ri tH_V}  u = Q^m u+\ri \int_0^t  e^{\ri sH_V}[H_V, Q^m]e^{-\ri sH_V}  u \, ds,
\end{equation}
where $[H_V, Q^m]$ is well-defined on $\CH^{m-1}(\Z^d)$. Indeed, noting that $[H_V, Q^m]=[Q^m,\Delta]$ for $H_V$ with any potential $V$, we have, for $\psi\in \CH^{m-1}(\Z^d)$ and $n\in\Z^d$,
\begin{eqnarray*}
([H_V, Q^m]\psi)_n&=&([Q^m,\Delta]\psi)_n\\
&=&\sum_{j=1}^d \left((1+|n|^2)^{\frac{m}2}-(1+|n+e_j|^2)^{\frac{m}2}\right)\psi_{n+e_j}\\
& &+ \, \sum_{j=1}^d \left((1+|n|^2)^{\frac{m}2}-(1+|n-e_j|^2)^{\frac{m}2}\right)\psi_{n-e_j},
\end{eqnarray*}
which implies that, there is some constant $\tilde c_m>0$, such that
$$\|[H_V, Q^m]\psi\|_0\leq 2\tilde c_m \|Q^{m-1}\psi\|_0=2\tilde c_m \|\psi\|_{m-1},$$
since, for $j=1,\cdots,d$, for $n\in\Z^d$,
\begin{eqnarray*}
\left|(1+|n|^2)^{\frac{m}2}-(1+|n\pm e_j|^2)^{\frac{m}2}\right|
&=&(1+|n|^2)^{\frac{m}2}\left|\left(1+\frac{2n_j\pm1}{1+|n|^2} \right)^{\frac{m}2}-1\right|  \\
&\leq&\frac{\tilde c_m}{d} (1+|n|^2)^{\frac{m-1}2}.\end{eqnarray*}
Hence, when $V$ is a real potential, we have, through \eqref{com-int-upper},
\begin{eqnarray*}
\|e^{-\ri tH_V}   u\|_{m}&=&\|e^{\ri tH_V}Q^{m} e^{-\ri tH_V}  u \|_0\\
&\leq& \|Q^{m} u\|_0+ \int_0^t \|[H_V, Q^{m}]e^{-\ri sH_V} u\|_0 \, ds\\
&\leq& \|u\|_m+ 2\tilde c_m\int_0^t \|e^{-\ri sH_V} u\|_{m-1} \, ds,\quad \forall \  t\geq 0.
\end{eqnarray*}
Since for $m=1$, $\|e^{-\ri tH_V}   u\|_{m-1}=\|u\|_{0}$, we obtain the linear upper bound for $\|e^{-\ri tH_V}   u\|_1$ for $u\in \CH^{1}(\Z^d)$:
$$\|e^{-\ri tH_V}   u\|_{1}\leq  \|u\|_1+2\tilde c_2\|u\|_0  t ,\quad \forall \  t\geq 0. $$
Then, via induction, for $u\in \CH^{2}(\Z^d)$,
\begin{eqnarray*}
\|e^{-\ri tH_V}   u\|_{2}&\leq& \|u\|_2+ \int_0^t \|e^{-\ri sH_V} u\|_1 \, ds\\
&\leq& \|u\|_2+ \|u\|_1 t  + \tilde c_2 \|u\|_0  t^2,\quad \forall \  t\geq 0.
\end{eqnarray*}
which shows \eqref{balistic-upper-2}. The above inequalities also imply that, for $m=1, 2$, there exists $c'_m>0$ such that
$$\|e^{-\ri tH_V}   u\|_{m}\leq   c'_m \|u\|_m (1+t)^m ,\quad \forall \  t\geq 0, $$
through which we obtain \eqref{balistic-upper-r} for $r=1,2$.

Assume that, for $m\in\N^*$, there exists a constant $c'_m>0$ such that, for any $u\in \CH^{m}(\Z^d)$,
%\marginpar{\clo We don't have this for for all $t > 0$.}
$$
\|e^{-\ri tH_V}   u\|_m\leq c'_m \|u\|_{m} (1+t)^m,\quad \forall \  t\geq 0.
$$
By induction, we have
%\marginpar{\clo Here we are actually using the bound for all $t > 0$.}
\begin{eqnarray*}
\|e^{-\ri tH_V}   u\|_{m+1}&\leq& \|u\|_{m+1}+ 2\tilde c_{m+1}\int_0^t \|e^{-\ri sH_V} u\|_{m} \, ds\\
&\leq& \|u\|_{m+1}+ \frac{2\tilde c_{m+1} c'_m}{m+1}  \|u\|_{m} (1+t)^{m+1},\quad \forall \  t\geq 0.
\end{eqnarray*}
Therefore, the upper bound \eqref{balistic-upper-r} is shown when $r>0$ is an integer.

For $r\in (m,m+1)$ with some $m\in\N$ ($ = \{0,1,2,\ldots\}$) we have
$$r=\left((m+1)-r\right)m+\left(r-m\right)(m+1).$$
%\marginpar{Do you mean the Riesz-Thorin interpolation theorem? If so, let's mention it, this makes it easier for the reader to look it up.}
Applying H\"older’s inequality, we obtain, for $\psi\in\CH^{m+1}(\Z^d)$,
\begin{eqnarray}
\|\psi \|^2_r&=&\sum_{n\in\Z^d}(1+|n|^2)^{\left((m+1)-r\right)m+\left(r-m\right)(m+1)}  \left|\psi_n\right|^{2}\nonumber\\
&\leq&\left(\sum_{n\in\Z^d}(1+|n|^2)^{m} \left|\psi_n\right|^2\right)^{(m+1)-r}
\left(\sum_{n\in\Z^d}(1+|n|^2)^{m+1} \left|\psi_n\right|^2\right)^{r-m}\nonumber\\
&=&\|\psi \|_m^{2(m+1-r)} \|\psi \|_{m+1}^{2(r-m)}.\label{interpol}
\end{eqnarray}
%\marginpar{Should $ \|u\|_0$ be replaced by $ \|u\|_{m+1}$ on the RHS of \eqref{interpo-upper}?}
In view of the integer order ballistic upper bound, for $u\in\CH^{m+1}(\Z^d)$,
\begin{eqnarray}
\|e^{-\ri t H_V} u \|_r&\leq& \|e^{-\ri t H_V} u \|_m^{m+1-r} \|e^{-\ri t H_V} u \|_{m+1}^{r-m}\nonumber\\
&\leq&  c^{m+1-r}_m c^{r-m}_{m+1} \|u \|_m^{m+1-r} \| u \|_{m+1}^{r-m}\cdot  t^{\left((m+1)-r\right)m+\left(r-m\right)(m+1)} \nonumber\\
&\leq&c_r \|u\|_r t^r ,  \quad  {\rm with} \   c_r:=  c^{m+1-r}_m c^{r-m}_{m+1} . \label{interpo-upper}
\end{eqnarray}
Note that $\CH^{m+1}(\Z^d)$ is dense in $\CH^r(\Z^d)$ in the sense that, if $u\in\CH^r(\Z^d)$, then for $\varepsilon>0$, there exists $u_\varepsilon\in\CH^{m+1}(\Z^d)$ such that
$\|u_\varepsilon-u\|_r< \varepsilon$.
Since \eqref{interpo-upper} implies that, for fixed $t\geq 0$, the time-evolution $e^{-\ri t H_V}: \CH^{m+1}(\Z^d)\to \CH^r(\Z^d)$ is bounded with operator norm controlled by $c_r t^r$.
By density and the continuous extension theorem for bounded linear operators
defined on a dense subspace, $e^{-\ri t H_V}$ extends uniquely to a bounded operator on $\CH^r(\Z^d)$ and (\ref{balistic-upper-r}) is satisfied.\qed

%Similar to (\ref{integal-second})
%\begin{eqnarray*}
%\|Q^r e^{-\ri tH}  \chi_I(H)  u\|_0^2&=&\la  \chi_I(H)  u, e^{\ri tH} Q^{2r} e^{-\ri tH}  \chi_I(H)   u \ra\nonumber \\
%&=& \la  \chi_I(H)  u,  Q^{2r}   \chi_I(H)   u \ra - \ri t \la \chi_I(H)  u, [H,- Q^{2r}]   \chi_I(H)   u  \ra\nonumber \\
% & & + \, \int_0^t \int_0^s   \left\la  e^{-\ri \tau H}  \chi_I(H)  u, [H,[H,-Q^{2r}]] e^{-\ri \tau H}  \chi_I(H)  u \right\ra \, d\tau \, ds.
%\end{eqnarray*}

\subsection{Ballistic Lower Bound}

Let us next focus on establishing the lower bound in \eqref{ballistic-decayingV} for $H_V$ with the potential $V$ satisfying \eqref{decayV-disc}.
For $r=1$, it is shown in Proposition \ref{prop-mourre-ballistic}. The order$-r$ ballistic lower bound for $r\geq 1$ is deduced from the following lemma.

\begin{lemma}\label{lemJensen}
For any self-adjoint operator $H$ on $\ell^2(\Z^d)$, if there exists $r>0$ such that for any non-vanishing $u\in \CH_{\rm ac}^r(\Z^d)$, $\|e^{-\ri tH} u \|_{r} \geq b_{r,u} t^r$ for some constant $b_{r,u} > 0$,
then for any $r'>r$ and $u\in \CH_{\rm ac}^{r'}(\Z^d)\setminus\{0\}$, there exists a constant $b_{r',u} > 0$ such that
$\|e^{-\ri tH} u \|_{r'} \geq b_{r',u} t^{r'}$.
\end{lemma}

\begin{remark}
This argument is a variation of the argument given in the proof of \cite[Lemma~2.7]{DT2010}.
\end{remark}

\proof Suppose that $u\in \CH_{\rm ac}^r(\Z^d)$ is  $\ell^2$-normalized. By assumption, there exists $b_{r,u} > 0$ such that $\| e^{-\ri tH} u \|_{r} \geq b_{r,u}  t^r$, which means that
$$
\sum_{n \in \Z^d} (1+|n|^2)^{r} |\la \delta_n , e^{-\ri tH} u \ra|^2  \geq b^2_{r,u}  t^{2r}.
$$
For $r' > r$, note that $x \mapsto |x|^{\frac{r'}{r}}$ is convex on $\R$. Using Jensen's inequality (and emphasizing that $u$ is $\ell^2$-normalized so that $n \mapsto |\langle \delta_n , e^{-\ri tH} u \rangle|^2$ is a probability distribution on $\Z^d$), we observe that, for any $\ell^2$-normalized $u\in \CH_{\rm ac}^{r'}(\Z^d)\subset  \CH_{\rm ac}^{r}(\Z^d)$,
\begin{align*}
\sum_{n \in \Z^d} (1+|n|^2)^{r'} |\langle \delta_n , e^{-\ri tH} u \rangle|^2 & = \sum_{n \in \Z^d} \left( (1+|n|^2)^{r} \right)^{\frac{r'}{r}} |\langle \delta_n , e^{-\ri tH} u \rangle|^2 \\
& \geq \left( \sum_{n \in \Z^d}(1+|n|^2)^{r}  |\langle \delta_n , e^{-\ri tH} u \rangle|^2 \right)^{\frac{r'}{r}}  \\
& \geq b^{\frac{2r'}{r}} _{r,u}  t^{2r'},
\end{align*}
which implies that $\| e^{-\ri tH} u \|_{r} \geq b_{r',u}  t^{r'}$ with $b_{r',u} =b_{r,u}^{\frac{r'}{r}} $. By scaling, the bound extends to non-normalized $u\in \CH_{\rm ac}^{r'}(\Z^d)\setminus\{0\}$.\qed

\medskip

As for the case $0<r<1$, we have the following lemma, which completes the proof of Theorem \ref{thm-ballistic-decay}.

\begin{lemma}\label{lemSmallorder}
For the Schr\"odinger operator $H_V$ with $V$ satisfying \eqref{decayV-disc}, for any $0<r<1$ and any non-vanishing $u\in \CH_{\rm ac}^r(\Z^d)$, there exists a constant $b_{r,u} > 0$ such that for $t \ge 1$,
$$
\|e^{-\ri tH_V} u \|_{r} \geq b_{r,u} t^{r}.
$$
\end{lemma}
\proof
Applying H\"older's inequality as (\ref{interpol}), we obtain, for $0<r<1$,
$$\|e^{-\ri t H_V}u\|_1\leq \|e^{-\ri t H_V}u\|_r^{\frac{1}{2-r}}\|e^{-\ri t H_V}u\|^{\frac{1-r}{2-r}}_2.$$
According to (\ref{balistic-upper-2}) in Proposition \ref{prop-upper} and Proposition \ref{prop-mourre-ballistic}, for any $u\in\CH^2_{\rm ac}(\Z^d)\setminus\{0\}$,
one can find $0<\alpha<2$ and $I_\alpha\subset[-2d,2d]\setminus\CZ$
such that $\chi_{I_{\alpha}}(H_V) u$ is non-vanishing and there exists $a_1>0$ such that
\begin{eqnarray*}
\|e^{-\ri t H_V}u\|_1 &\geq& a_1 \alpha^\frac12  \|\chi_{I_\alpha}(H_V) u\|_{0} \, t, \\
\|e^{-\ri t H_V}u\|_2&\leq& \|u\|_2+ \|u\|_1 \, t+ \tilde c_2  \|u\|_0 \, t^2,
\end{eqnarray*}
Hence, as $t\to\infty$,
$$
\|e^{-\ri t H_V}u\|_r\geq \frac{ (a_1 \alpha^\frac12 \|\chi_{I_{\alpha}}(H_V) u\|_{0} t)^{2-r}}{(\|u\|_2+ \|u\|_1 t+ \tilde c_2  \|u\|_0 t^2)^{1-r}}
\geq \frac{(a_1 \alpha^\frac12 \|\chi_{I_{\alpha}}(H_V) u\|_{0})^{2-r}}{2^{1-r}(\tilde c_2  \|u\|_0 )^{1-r}} t^r.
$$

Note that $\CH^{2}_{\rm ac}(\Z^d)=\CH^{2}(\Z^d)\cap \CH_{\rm ac}(H_V)$ is dense in $\CH^r_{\rm ac}(\Z^d)=\CH^r(\Z^d)\cap \CH_{\rm ac}(H_V)$
%\marginpar{Should we write ``$\CH^{2}_{\rm ac}(\Z^d)$ is dense in $\CH^r_{\rm ac}(\Z^d)$''?}
in the sense that, given $u\in\CH_{\rm ac}^r(\Z^d)$, for any $\varepsilon>0$, there exists $u_\varepsilon\in \CH_{\rm ac}^{2}(\Z^d)$ such that
 $\|u_\varepsilon-u\|_{r}< \varepsilon$, which implies that, for $\varepsilon$ sufficiently small,
 \begin{eqnarray*}\|\chi_{I_{\alpha}}(H_V) u_\varepsilon\|_0&\geq&\|\chi_{I_{\alpha}}(H_V) u\|_0- \|u_\varepsilon-u\|_0\\
&\geq&\|\chi_{I_{\alpha}}(H_V) u\|_0- \|u_\varepsilon-u\|_r  \  \geq \ \frac12\|\chi_{I_{\alpha}}(H_V) u\|_0,\\
 \|u_\varepsilon\|_0&\leq&   \|u\|_0 +  \|u_\varepsilon-u\|_0  \  \leq   \   2\|u\|_0.
\end{eqnarray*}
According to (\ref{balistic-upper-r}) in Proposition \ref{prop-upper}, for $t\geq 0$,
$$\|e^{-\ri t H_V}(u_\varepsilon-u)\|_{r}\leq c_r \|u_\varepsilon-u\|_{r}  t^r.$$
 Then, for $\varepsilon$ sufficiently small, as $t\to\infty$, we have
 %\marginpar{Here it is important to make the norm on the RHS explicit.}
\begin{align*}
\|e^{-\ri t H_V}u\|_r & \geq \|e^{-\ri t H_V}u_\varepsilon\|_r -  \|e^{-\ri t H_V}(u_\varepsilon-u)\|_{r}\\
& \geq \frac{ (a_1 \alpha^\frac12 \|\chi_{I_{\alpha}}(H_V) u_\varepsilon\|_{0})^{2-r}}{2^{1-r}(\tilde c_2  \|u_\varepsilon\|_0 )^{1-r}} \cdot t^r -c_r \|u_\varepsilon-u\|_{r}  t^r \\
& \geq \left(\frac{(a_1 \alpha^\frac12 \|\chi_{I_{\alpha}}(H_V) u\|_{0})^{2-r}}{2^{4-3r} (\tilde c_2  \|u\|_0 )^{1-r}}  -c_r\varepsilon \right)t^r \\
& \geq  \frac{(a_1 \alpha^\frac12 \|\chi_{I_{\alpha}}(H_V) u\|_{0})^{2-r}}{16 (\tilde c_2  \|u\|_0 )^{1-r}} t^r.\qed
%\quad {\rm with}  \  b_r:=\frac{(a_1 \theta^\frac12 \|\chi_{I_{\alpha}}(H_V) u\|_{0})^{2-r}}{2^{2-r} (\tilde c_2  \|u\|_0 )^{1-r}}
\end{align*}

\section*{Acknowledgment}

We are grateful to Sergey Denisov, Jake Fillman, and Tal Malinovitch for helpful information about the literature and to Marc-Adrien Mandich for encouraging us to clarify in detail the absence of singular continuous spectrum assuming only \eqref{decayV-disc}.

\end{document}